\documentclass[twocolumn]{aastex62}

\graphicspath{{./plots/}}

\newcommand\ddfrac[2]{\frac{\displaystyle #1}{\displaystyle #2}}
\newcommand{\np}{star-forming}

\newcommand{\kapteyn}{Kapteyn Astronomical Institute, University of Groningen, P.O. Box 800, 9700 AV, Groningen, The Netherlands}
\newcommand{\dawn}{The Cosmic Dawn Center, Niels Bohr Institute, University of Copenhagen, Juliane Maries Vej 30, DK-2100 Copenhagen {\O}, Denmark}
\newcommand{\harvard}{Center for Astrophysics | Harvard \& Smithsonian, Optical and Infrared Astronomy Division, 60 Garden St., Cambridge, MA 02138, USA}
\newcommand{\lam}{Aix Marseille Universit\'e, CNRS, LAM (Laboratoire d'Astrophysique de Marseille) UMR 7326, F-13388, Marseille, France}
\newcommand{\dark}{Dark Cosmology Centre, Niels Bohr Institute, University of Copenhagen, Juliane Maries Vej 30, DK-2100 Copenhagen {\O}, Denmark}

\received{XXX}
\revised{YYY}
\accepted{ZZZ}

\submitjournal{ApJ}

\shorttitle{The SHMRs of passive and star-forming galaxies in SMUVS}
\shortauthors{Cowley, W.I. et al.}

\begin{document}

\title{\uppercase{The Stellar-to-Halo Mass Ratios of Passive and Star-Forming Galaxies at $\lowercase{z}\sim2-3$ from the SMUVS Survey}}

\correspondingauthor{William~{I.}~Cowley}
\email{cowley@astro.rug.nl}

\author[0000-0003-1314-2641]{William~{I.}~Cowley}
\affil{\kapteyn}
\author[0000-0001-8183-1460]{Karina~{I.}~Caputi}
\affil{\kapteyn}
\affil{\dawn}
\author[0000-0001-7264-6925]{Smaran~Deshmukh}
\affil{\kapteyn}
\author[0000-0002-3993-0745]{Matthew~{L.}~{N.}~Ashby}
\affil{\harvard}
\author[0000-0002-0670-0708]{Giovanni~{G.}~Fazio}
\affil{\harvard}
\author[0000-0001-5891-2596]{Olivier~Le~F\`evre}
\affil{\lam}
\author[0000-0002-8149-8298]{Johan~{P.}~{U.}~Fynbo}
\affil{\dark}
\affil{\dawn}
\author[0000-0002-7303-4397]{Oliver~Ilbert}
\affil{\lam}
\author[0000-0002-2281-2785]{Bo~Milvang-Jensen}
\affil{\dark}
\affil{\dawn}

\begin{abstract}
In this work, we use measurements of galaxy stellar mass and two-point angular correlation functions to constrain the stellar-to-halo mass ratios (SHMRs) of passive and \np\ galaxies at $z\sim2-3$, as identified in the \emph{Spitzer} Matching Survey of the UltraVISTA ultra-deep Stripes (SMUVS). We adopt a sophisticated halo modeling approach to statistically divide our two populations into central and satellite galaxies. For central galaxies, we find that the normalization of the SHMR is greater for our passive population. Through the modeling of $\Lambda$ cold dark matter halo mass accretion histories, we show that this can only arise if the conversion of baryons into stars was more efficient at higher redshifts and additionally that passive galaxies can be plausibly explained as residing in halos with the highest formation redshifts (i.e., those with the lowest accretion rates) at a given halo mass. At a fixed stellar mass, satellite galaxies occupy host halos with a greater mass than central galaxies, and we find further that the fraction of passive galaxies that are satellites is higher than for the combined population. This, and our derived satellite quenching timescales, combined with earlier estimates from the literature, support dynamical/environmental mechanisms as the dominant process for satellite quenching at $z\lesssim3$.
\end{abstract}


\section{Introduction} 
\label{sec:intro}
One of the most striking properties of the galaxy population at $z=0$ is its distinct optical color bi-modality \citep[e.g.,][]{Strateva:2001, Baldry:2004}. Broadly speaking, galaxies are either blue or red, commonly interpreted as star-forming or passively evolving (meaning the current ongoing star formation is significantly lower than the past average). respectively, with a transitional `green' valley \citep[e.g.,][]{Schawinski:2014} in between. This bi-modality is observed to be in place until at least $z\sim1$ \citep[e.g.,][]{Bell:2004} and possibly persists at higher redshifts \citep[e.g.,][]{Cirasuolo:2007}. Although the competing effect of dust obscuration becoming increasingly important at higher redshifts makes the interpretation of the optical color less clear \citep[e.g.,][]{Stern:2006}. 

The relative importance of various physical processes through which this bi-modality emerges is not immediately clear. However, over the last few years, a broad paradigm of two main channels for quenching star formation has emerged. These can be summarized as `mass' (sometimes referred to as `intrinsic') and `environment' quenching and have been plausibly shown to explain the observed Schechter function shape of the local galaxy stellar mass function \citep{Peng:2010}. We discuss these two main channels in greater detail in the following two paragraphs.           

For mass/intrinsic quenching, some early galaxy formation models produced a distinct red sequence through the incorporation of radio-mode active galactic nucleus (AGN) feedback \citep[e.g.][]{Bower:2006, Croton:2006} where star formation is suppressed in relatively high-mass halos ($\gtrsim10^{12}$~M$_{\odot}$) in which gas in the hot halo is undergoing quasi-static cooling. However, the bi-modality in the model of \cite{Menci:2005} was produced by the mass assembly history of dark matter halos, as their model did not include a prescription for AGN feedback. It should also be noted that whilst physical models can produce a distinct red sequence, they often struggle to precisely reproduce the observed distribution of $g-r$ color locally \citep[e.g.,][]{Trayford:2015, Lacey:2016}.

It is also accepted that environmental effects acting dynamically on satellite galaxies orbiting within the virial radius of a host halo also play a role in producing a population of passive galaxies. These environmental mechanisms include ram pressure stripping of a galaxy's hot gas reservoir as it enters that of a larger host halo, strangulation as gas no longer cools onto the satellite galaxy and harassment, tidal interactions between other nearby galaxies \citep[e.g.][]{vdBosch:2008}. Though, as we will discuss later, outflows from supernovae in the satellite may also be important \citep{McGee:2014}. 

In this work, we use a halo occupation formalism \citep[e.g.][]{Zheng:2005} to investigate the differences between passive and \np\ galaxies at $z=2-3$ as identified in the \emph{Spitzer} Matching survey of the UltraVISTA ultra-deep Stripes \citep[SMUVS;][see also \citealt{Caputi:2017}]{Ashby:2018, Deshmukh:2018}. This halo formalism provides a statistical framework that describes the expected mean number of galaxies occupying host halos of a given mass. More specifically, we use a halo model similar to that introduced by Leauthaud et al. (\citeyear{Leauthaud:2011}, see also \citealt{Tinker:2013}), which begins with a parameterization of the stellar-to-halo mass ratio (SHMR) for central galaxies, a function that specifies the mean stellar mass of a central galaxy as a function of halo mass. The free parameters in this model are then calibrated using the observed galaxy stellar mass and two-point angular correlation functions. Although this is a statistical model that lacks a direct physical interpretation, it provides us with a means to investigate the link between stellar and halo mass for passive and \np\ galaxies (divided into central and satellite galaxies). As such, the model's predictions can then be understood in the context of the potential physical processes involved.        

We restrict our analysis to $z=2-3$ as the SMUVS photometry is optimized for $z\gtrsim2$ galaxies \citep{Deshmukh:2018}. At higher redshifts ($z\gtrsim3$), the number of passive galaxies is insufficient to robustly measure the angular two-point correlation function for this population, which is a key constraint on our model. We thus extend similar analyses that have been performed at lower redshifts using a combination of galaxy-galaxy lensing, galaxy two-point correlation functions, stellar mass functions and halo model analyses \citep[e.g.][]{Mandelbaum:2006, vanUitert:2011, Tinker:2013, RodriguezPuebla:2015}. These earlier works generally find that local ($z\sim0$) differences in the SHMR between passive and star-forming galaxies are generally small, with blue/star-forming galaxies having an SHMR a factor of $\sim2$ higher than red/passive galaxies \citep[e.g.][]{RodriguezPuebla:2015}. However, this difference diminishes toward higher redshift and  actually inverts such that at a fixed stellar mass, star-forming galaxies are in \emph{more} massive halos by $z\sim1$ \citep[e.g.][]{Tinker:2013}, particularly for galaxies with high stellar masses ($\gtrsim10^{10.5}$~M$_{\odot}$).  

Throughout this work, we assume a flat $\Lambda$ cold dark matter (CDM) cosmological model with ($\Omega_{\rm m}$, $\Omega_{\rm b}$, $\Omega_{\Lambda}$, $h$, $\sigma_{8}$, $n_{\rm s}$) = ($0.3$, $0.04$, $0.7$, $0.8$, $0.95$). This paper is structured as follows: in Section~\ref{sec:data}, we summarize the main details of the SMUVS survey, describe the data used and the determination of our observational constraints, the galaxy stellar mass, and angular two-point correlation functions. In Section~\ref{sec:model}, we describe our halo occupation model, and we present our main results in Section~\ref{sec:results}. We summarize and conclude in Section~\ref{sec:conclude}. Some further information regarding our model fitting is given Appendix~\ref{sec:fitting_appendix}.   
\section{The Data}
\label{sec:data}
\subsection{The SMUVS survey}
The SMUVS program (PI: K. Caputi; Ashby et al. \citeyear{Ashby:2018}) has collected ultra-deep \emph{Spitzer} $3.6$ and $4.5$~$\mu$m data within the COSMOS\footnote{\url{http://cosmos.astro.caltech.edu}} field \citep{Scoville:2007} covering three of the UltraVISTA ultra-deep stripes \citep{McCracken:2012} with deep optical coverage from the Subaru Telescope \citep{Taniguchi:2007}. The UltraVISTA data considered here correspond to the third data release,\footnote{\url{http://www.eso.org/sci/observing/phase3/data_releases/uvista_dr3.pdf}} which reaches an average depth of $K_{\rm s}=24.9\pm0.1$ and $H=25.1\pm0.1$ ($2$~arcsec diameter, $5\sigma$).   

A thorough description of the SMUVS multiwavelength source catalog construction and spectral energy distribution (SED) fitting is given in \cite{Deshmukh:2018}; however, we summarize the main details here.

Sources are extracted from the UltraVISTA $HK_{\rm s}$ average stack mosaics using {\sc SExtractor} \citep{BertinArnouts:1996}. The positions of these sources are then used as priors to perform iterative point-spread function (PSF) fitting photometric measurements on the SMUVS $3.6$ and $4.5$~$\mu$m mosaics, using the {\sc daophot} package \citep{Stetson:1987}.

For all of these sources, $2$~arcsec diameter circular photometry on $26$ broad, intermediate, and narrow bands ($U$ to $K_{\rm s}$) is measured \citep{Deshmukh:2018}. After cleaning for Galactic stars using a $B$-$J$-$[3.6]$ color selection \citep[e.g.,][]{Caputi:2011}, and masking regions of contaminated light around the brightest sources, the final catalog contains $\sim2.9\times10^5$ UltraVISTA sources with a detection in at least one IRAC band over an area of $\sim0.66$ square degrees.

The SED fitting is performed with all $28$ bands ($26$ $U$ through $K_{\rm s}$ as well as \emph{Spitzer} $3.6$ and $4.5$~$\mu$m) using the template-fitting code {\sc LePhare}\footnote{\url{http://www.cfht.hawaii.edu/~arnouts/LEPHARE/lephare.html}} \citep{Arnouts:1999,Ilbert:2006}. We assume \cite{BruzualCharlot:2003} templates corresponding to a simple stellar population formed with a \cite{Chabrier:2003} stellar initial mass function and either solar or sub-solar ($1$~Z$_{\odot}$ or $0.2$~Z$_{\odot}$) metallicity, and allow for the addition of nebular emission lines. Additionally, we assume exponentially declining star formation histories. 

Photometric redshifts and stellar mass estimates are obtained for $>99$~\% of our sources. Using ancillary spectroscopic data in COSMOS to assess the quality of the obtained photometric redshifts, we found that the standard deviation, $\sigma_{z}$, of $|z_{\rm phot}-z_{\rm spec}| / (1 + z_{\rm spec})$, based on $\sim1.4\times10^4$ galaxies with reliable spectroscopic redshifts in the COSMOS field (see Table~1 in Ilbert et al. \citeyear{Ilbert:2013} and references therein) is $0.026$ \citep{Deshmukh:2018} and is $0.035$ for the $z\sim2-3$ population. This statistic is computed excluding outliers, defined as objects for which $|z_{\rm phot}-z_{\rm spec}| / (1 + z_{\rm spec})>0.15$, which comprise only $\sim5.5$~\% of the whole spectroscopic catalog. These results compare favorably with other photometric surveys in the literature \citep[e.g.][]{Ilbert:2013,Laigle:2016} and highlight the high accuracy of our derived photometric redshifts. For our parent sample of $z\sim2-3$ galaxies, we find $\sigma_{z}=0.035$.       

Throughout this paper, we use the best-fit redshifts and stellar masses computed by {\sc LePhare}. We also limit ourselves to $\log(M_{\star}/\mathrm{M}_{\odot})>9.8$ to ensure a high level of completeness for both our passive and \np\ galaxies. Our combined (i.e., passive + \np) galaxy population is $95$~\% complete for $\log(M_{\star}/\mathrm{M}_{\odot})>9.6$ at $z\sim2-3$ \citep{Deshmukh:2018}. This stellar mass cut leaves us with $\sim10^{3}$ passive and $\sim6.8\times10^{3}$ star-forming galaxies with $2<z<3$.
\subsection{Definition of passive galaxies}

Passive galaxies, i.e., those for which the current star formation is negligible relative to their past average, are defined using the criteria of \cite{Deshmukh:2018}. A galaxy is defined as passive if: (i) it has a color excess E$(B-V)\leq0.1$ (indicating that interstellar dust has little impact on the SED) as computed by {\sc LePhare}; \emph{and} (ii) it has a rest-frame color $u-r\geq1.3$ i.e., is `red'. The use of the color cut in conjunction with the SED extinction prevents dust-obscured star formation from being the main cause of the red color. This leaves an old stellar population with negligible ongoing star formation as the most plausible explanation of the galaxy's SED.       

\subsection{The galaxy stellar mass function}
We compute the galaxy stellar mass functions for our passive and \np\ populations using the $1/V_{\mathrm{max}}$ technique \citep{Schmidt:1968} as was done by \cite{Deshmukh:2018}. A $V_{\mathrm{max}}$ correction is computed for each galaxy based on the maximum volume at which it would have a magnitude $[4.5$~$\mu\mathrm{m}]=26$ (or $[3.6$~$\mu\mathrm{m}]=26$ in the case of a non-detection at 4.5~$\mu$m). For fainter sources, no $V_{\mathrm{max}}$ correction is applied. A further completeness correction is applied to each galaxy based on its $4.5$~$\mu$m magnitude (or $3.6$~$\mu$m in the case of a non-detection at $4.5$~$\mu$m) and the ratio of SMUVS number counts to the completeness corrected number counts derived from the \emph{Spitzer}-CANDELS/COSMOS survey \citep{Ashby:2015} at that magnitude.

The errors are derived by summing a Poisson noise term, $\sigma_{\mathrm{Poi}}$, an error associated with photometric uncertainties, $\sigma_{\mathrm{mc}}$, and a cosmic variance error, $\sigma_{\mathrm{cv}}$, in quadrature. The Poisson term, $\sigma_{\rm Poi}$, is derived using the tabulated values of \cite{Gehrels:1986}. The photometric term, $\sigma_{\rm mc}$, is derived by scattering the photometry input to {\sc LePhare} within its uncertainties for each galaxy to create $100$ realizations of the SMUVS catalog and recomputing the stellar mass function for each; $\sigma_{\rm mc}$ is then taken to be the $16-84^{\rm th}$ percentile of these mock catalog stellar mass functions in each stellar mass bin. The cosmic variance term is derived using the prescription of \cite{Moster:2011}.

Deshmukh et al. (\citeyear{Deshmukh:2018}, see their Fig.~19) provides a thorough comparison of $2<z<6$ SMUVS-derived galaxy stellar mass functions to earlier estimates in the literature.  
  
\subsection{The two-point angular correlation function}
The two-point angular autocorrelation function, $w(\theta)$, describes the excess probability, compared to a random (Poisson) distribution, of finding a pair of galaxies at some angular separation $\theta > 0$. It is defined such that
\begin{equation}
\delta^{2}P_{12} = \bar{\eta}^{2}\,[1 + w(\theta)]\,\delta\Omega_{1}\,\delta\Omega_{2}\rm,
\end{equation}
where $\bar{\eta}$ is the mean surface density of the population per unit solid angle, $\theta$, is the angular separation, and $\delta\Omega_{\rm i}$ is a solid angle element.
Here, the angular correlation function is computed according to the standard \cite{LandySzalay:1993} estimator
\begin{equation}
w(\theta) = \frac{DD(\theta) - 2DR(\theta) + RR(\theta)}{RR(\theta)}\rm,
\label{eq:wtheta}
\end{equation}
where $DD$, $DR$ and $RR$ represent the number of data-data, data-random, and random-random pairs in a bin of angular separation, respectively. The random catalog is constructed to have the same angular selection as the data by generating random points over the survey area and discarding those that fall within masked areas relating to, e.g., foreground stars. Due to the high stellar mass completeness of our samples ($\gtrsim95$~\%) we do not need to consider any variations in sensitivity/depth across our survey. We use a single random catalog with $\sim5\times10^5$ objects ($\sim50\times$ greater than our largest galaxy sample) throughout our analyses. The errors on the two-point angular correlation function are estimated from the data using a jackknife method \citep{Norberg:2009}. We divide the SMUVS footprint into $60$ approximately equal area regions. Removing one region at a time, we compute the covariance matrix as 
\begin{equation}
C_{i,j} = \frac{N-1}{N}\sum_{l=1}^{N}(w_{i}^{l}-\bar{w_{i}})\times(w_{j}^{l}-\bar{w_{j}})\rm,
\label{eq:covar_matrix}
\end{equation}
where $N$ is the total number of regions, $\bar{w}$ is the mean correlation function ($\sum_{l}w^{l}/N$) and $w^{l}$ is the estimate of $w$ with the $l^{\rm th}$ region removed (i.e. computed from 59 of the 60 regions).  We compute $w(\theta)$ using eight evenly spaced logarithmic bins of angular separation in the range $-3<\log_{10}(\theta/\mathrm{deg})<-1$.

\section{The Theoretical Model}
\label{sec:model}
The model used here is based on that presented by \cite{Tinker:2013}, which in turn extends on that first presented by \cite{Leauthaud:2011}. We begin by defining the mean stellar-to-halo mass relation, $f_{\rm SHMR}$, for central galaxies. For this we adopt 
\begin{equation}
\log\,f_{\rm SHMR} = \log[M_{\star}(M_{\mathrm{h}})] = \log[\varepsilon(M_{\rm h})\times M_{\rm h}]
\label{eq:fshmr}
\end{equation} 
where the SHMR, $\varepsilon(M_{\rm h})$, is described by
\begin{equation}
\varepsilon(M_{\rm h}) = 2\epsilon_{0}\left[\left(\frac{M_{\rm h}}{M_{\rm c}}\right)^{-\beta} + \left(\frac{M_{\rm h}}{M_{\rm c}}\right)^{\gamma}\right]^{-1}
\label{eq:fshmr_epsilon}
\end{equation}

That is, a double power law with a low-mass slope, $\beta$, and a high-mass slope, $-\gamma$, normalized to $\epsilon_{0}$ at halo mass $M_{\rm c}$ \citep[e.g.,][]{Yang:2003,Moster:2010}. More complex forms for this relation have been suggested in the literature \citep[e.g.,][]{Behroozi:2013b}; however, we find that a double power law is sufficient for our purposes. We do not place any constraint on $f_{\rm SHMR}$ other than the requirement that it is monotonic with halo mass such that it can be inverted, which in practice mainly limits the possible values of $\gamma$. We allow for an intrinsic Gaussian scatter from this relation, $\sigma_{\log M_{\star\rm,int}}$, of $0.16$~dex, which we assume is constant with respect to halo mass \citep[e.g.,][]{More:2011}. We also assume a measurement-based Gaussian scatter of $\sigma_{\log M_{\star\rm,meas}}=0.04\times(1+z)$ as suggested by \cite{Ilbert:2013}, which is $0.13$~dex based on a median redshift of $z=2.3$ for our $\log(M_{\star}/M_{\odot})>9.8$ galaxies. We thus use a total scatter of $\sigma_{\log M_{\star}}=\left[\sigma_{\log M_{\star\rm,int}}^{2}+\sigma_{\log M_{\star\rm,meas}}^{2}\right]^{1/2}\approx0.2$~dex. For simplicity, we apply this fixed value to both passive and \np\ galaxies.  

At each halo mass, we assume that a fraction, $f_{\rm p}$, of our central galaxies are passive. For this we assume a smooth hyperbolic tan function
\begin{equation}
f_{\rm p}(M_{\rm h}) = f_{\rm max,p}\left[\tanh\left(\alpha_{\rm p}[\log(M_{\rm h}) - \log(M_{\rm h,p})]\right) + 1\right] / 2\rm,
\label{eq:frac_cen_p}
\end{equation}
that transitions between having zero and a fraction of $f_{\rm max,p}$ passive central galaxies, with a fraction of $f_{\rm max,p}/2$ galaxies quenched at $M_{\rm h,p}$ and the width of the transition being controlled by $\alpha_{\rm p}$. The choice of this functional form is somewhat ad hoc and exponential functional forms have also been used elsewhere in the literature \cite[e.g.][]{TinkerWetzel:2010}, though we find that implementing these here have a minimal impact on our results.

The halo occupation distribution (HOD) function of central galaxies is described by 

\begin{equation}
N_{\rm c}(M_{\rm h}| > M_{\star})=\frac{1}{2}\left[1 - \mathrm{erf}\left(\frac{\log M_{\star} - \log\,f_{\rm SHMR}}{2\sigma_{\log M_{\star}}} \right)\right]\rm,
\end{equation}  

\citep{Leauthaud:2011}. For passive central galaxies, this is multiplied by $f_{\rm p}(M_{\rm h})$ and by $1-f_{\rm p}(M_{\rm h})$ for \np\ centrals.

The occupation of satellite galaxies is modeled as a power law with an exponential cut-off,

\begin{equation}
N_{\rm s}(M_{\rm h}| > M_{\star}) = \left(\frac{M_{\rm h}}{M_{\rm sat}}\right)^{\alpha_{\rm sat}}\exp\left(-\frac{M_{\rm cut}}{M_{\rm h}}\right)\rm,
\label{eq:nsat}
\end{equation} 
\citep{Leauthaud:2011} where $\alpha_{\rm sat}$ is the power-law slope the characteristic satellite halo mass, $M_{\rm sat}$, is described by
\begin{equation}
\frac{M_{\rm sat}}{10^{12}\mathrm{M}_{\odot}} = B_{\rm sat}\left(\frac{f_{\rm SHMR}^{-1}}{10^{12}\mathrm{M}_{\odot}}\right)^{\beta_{\rm sat}}\rm.
\label{eq:Msat}
\end{equation}
Here, $f_{\rm SHMR}^{-1}$ is the inverse of \autoref{eq:fshmr} and $B_{\rm sat}$ and $\beta_{\rm sat}$ describe the normalization and slope of the power law, respectively. The parameter $M_{\rm cut}$ in \autoref{eq:nsat} is often modeled in a similar fashion. However, for simplicity, here we fix this parameter according to
\begin{equation}
\log M_{\rm cut} = 0.76\log M_{\rm sat} + 2.3 \rm,
\end{equation}
following \cite{Conroy:2006}. Our combined HOD is the sum of the central and satellite distributions, i.e.,
\begin{equation}
N(M_{\rm h}| > M_{\star}) = N_{\rm c}(M_{\rm h}| > M_{\star}) + N_{\rm s}(M_{\rm h}| > M_{\star})\rm.
\label{eq:hod}
\end{equation}
\subsection{The stellar mass function}
From the HOD (\autoref{eq:hod}), we can calculate the stellar mass function $\Phi(M_{\star,1},M_{\star,2})$, i.e., the abundance of galaxies with $M_{\star,1}<M_{\star}<M_{\star,2}$, using
\begin{eqnarray}
\Phi(M_{\star,1},M_{\star,2})&=&\int_{0}^{\infty}[N(M_{\rm h}| > M_{\star,1})-N(M_{\rm h}| > M_{\star,2})]\nonumber\\&\times&\frac{\mathrm{d}n}{\mathrm{d}M_{\rm h}}\,\mathrm{d}M_{\rm h}\rm,
\label{eq:cond_gsmf}
\end{eqnarray}
where $\mathrm{d}n/\mathrm{d}M_{\rm h}$ is the halo mass function, from which we can construct the total stellar mass function.  This is then convolved with a Gaussian with $\sigma=\sigma_{\log M_{\star}}$ to account for scatter before comparing to our observational data.
\subsection{The two-point angular correlation function and the effect of photometric redshift errors}
\label{sec:project_photz}
As a further constraint on our model, we use the measured two-point angular correlation functions of passive and \np\ galaxies selected by their stellar mass. Using the angular correlation function, $w(\theta)$, rather than its spatial analog, $\xi(r)$, is a necessary consequence of our photometric redshifts, which are not accurate enough to measure the galaxy distribution in three dimensions. Therefore, a two-point spatial correlation function, $\xi(r)$, computed from an HOD, needs to be projected along the line of sight into two dimensions for comparison with our data. 

First, we compute $\xi(r)$ from our HOD assuming Navarro-Frenk-White \citep[NFW; ][]{NFW:1997} halo density profile, with the concentration relation of \cite{Bullock:2001}.\footnote{\cite{Martinez-Manso:2015} found that the predictions of halo models, such as the one used here, are fairly insensitive to the adopted concentration relation (see their Appendix~D).} The halo mass function used is the parameterization of \cite{Tinker:2008}, with the high-redshift correction of \cite{Behroozi:2013b}, as well as the large-scale dark matter halo bias parameterization of \cite{Tinker:2010}. These Tinker et al. relations adopt the definition of a halo as a spherical overdensity of $200$ relative to the mean cosmic density at the epoch of interest \citep[e.g.][]{LaceyCole:1994, Tinker:2008}.

The matter power spectrum is computed according to \cite{EisensteinHu:1995} with the nonlinear correction of \cite{Smith:2003} applied. Additionally, we implement the two-halo exclusion model of \cite{Tinker:2005}, which improves on that presented in \cite{Zheng:2004}.  

For projecting our two-point spatial correlation functions, we use the \cite{Limber:1953} equation,
\begin{equation}
w(\theta) = \ddfrac{\int \left(n_{\rm gal}(z)\,\frac{\mathrm{d}V}{\mathrm{d}z}\,W(z)\right)^{2}\frac{\mathrm{d}z}{\mathrm{d}\chi_{\rm c}}\,\mathrm{d}z\int\xi(r,z)\,\mathrm{d}u}{\left(\int n_{\rm gal}(z)\,\frac{\mathrm{d}V}{\mathrm{d}z}\,W(z)\,\mathrm{d}z\right)^{2}}\rm,
\label{eq:limber}
\end{equation}
where $n_{\rm gal}$ is the number density of galaxies predicted by the HOD, $\mathrm{d}V/\mathrm{d}z$ is the comoving volume element, $\mathrm{d}z/\mathrm{d}\chi_{\rm c}=H_{0}E(z)/c$ where $E(z)=[\Omega_{\mathrm{m}}(1+z)^{3} + \Omega_{\Lambda}]^{1/2}$, and $\chi_{\rm c}$ corresponds to the comoving radial distance to redshift $z$.  The comoving line-of-sight separation, $u$, is defined by $r=[u^{2} + \chi_{\rm c}^{2}\varpi^{2}]^{1/2}$ where $\varpi^{2}/2 = [1 - \cos(\theta)]$. 

The $W(z)$ term in \autoref{eq:limber} was introduced by \cite{Cowley:2018} and relates to the redshift window that is being probed by the survey. If the redshifts were known precisely, then (ignoring further complications such as redshift space distortions) this would be a top-hat function equal to unity between the limits of the redshift range probed and zero elsewhere. However, with photometric redshifts, this is not the case, and the top-hat window should be convolved with an error kernel that is generally unknown \emph{a priori}.

In order to mitigate this, here we assume a Gaussian error kernel and approximate a $(1+z)$ evolution in the error kernel dispersion, $\Delta_{z}$, such that
\begin{equation}
W(z) = \frac{1}{2}\left[\mathrm{erf}\left(\frac{z-z_{\rm lo}}{\Delta_{z}\,(1+z_{\rm lo})}\right) - \mathrm{erf}\left(\frac{z-z_{\rm hi}}{\Delta_{z}\,(1+z_{\rm hi})}\right) \right]\rm.
\label{eq:sigma_z_window}
\end{equation}
Here, $z_{\rm lo}$ and $z_{\rm hi}$ represent the lower and upper redshift limits of the photometric redshift bin, respectively. Whilst the integral in \autoref{eq:limber} is, in principle, over all redshift, it only has significant contributions from redshifts where $W(z)$ is appreciably nonzero.

Once projected according to \autoref{eq:limber}, we account for the integral constraint \citep[e.g.,][]{GrothPeebles:1977}. This is the correction required as the measured angular correlation function will integrate to zero over the whole field, by construction of the estimator for $w(\theta)$ used. This results in the measured angular correlation function being underestimated by an average amount
\begin{equation}
\sigma_{\rm IC}^{2} = \frac{1}{\Omega^{2}}\int\int w_{\rm true}(\theta)\,\mathrm{d}\Omega_{1}\mathrm{d}\Omega_{2}\rm,
\label{eq:integral_constraint}
\end{equation}
where $w_{\rm true}$ is the true angular correlation function and the angular integrations are performed over a field of area $\Omega$.

Here, we evaluate \autoref{eq:integral_constraint} according to the numerical method proposed by \cite{RocheEales:1999},
\begin{equation}
\sigma_{\rm IC}^{2} = \frac{\sum w_{\rm true}(\theta)\,RR(\theta)}{\sum RR(\theta)}\rm.
\end{equation}
We take $w_{\rm true}$ to be the angular correlation function predicted by our HOD model and subtract $\sigma_{\rm IC}^{2}$ from it before comparing it to our observed correlation functions, rather than subtract a value from our observed correlation functions. We do this for each evaluation of $w(\theta)$ in our fitting procedure, which is described below. These corrections are typically $\sim10^{-2}$.

\subsection{Fitting}

Our model has a total of $15$ free parameters. Three of these ($f_{\rm max,p}$, $\alpha_{\rm p}$, $M_{\rm h,p}$) relate to $f_{\rm p}(M_{\rm h})$ (\autoref{eq:frac_cen_p}). A further six for each population (passive and \np) come from the four $f_{\rm SHMR}$ parameters (\autoref{eq:fshmr}) ($\epsilon_{0}$, $M_{\rm c}$, $\beta$, $\gamma$) and two for satellite galaxies ($B_{\rm sat}$, $\beta_{\rm sat}$). We fix $\alpha_{\rm sat}=1$ \citep[e.g.,][]{Kravtsov:2004}, $\sigma_{\log M_{\star}}=0.2$ as discussed above and $\Delta_{z}=0.035$ (based on a comparison to our spectroscopic redshift sample) for both populations. In future, we expect to investigate more complex models for $\Delta_{z}$ (e.g., a dependence on stellar mass); however, we find that our data are unable to robustly constrain these, so we have opted for the simplest model and have fixed $\Delta_{z}$ to be constant here.    

In our fitting, we calibrate our model parameters such that we minimize the $\chi^{2}$, which we define as
\begin{eqnarray}
\chi^{2} &=& \sum_{\rm p,sf}\left[\sum_{l}\frac{\log\Phi^{\rm o}_{l} - \log\Phi^{\rm m}_{l}}{\log\sigma_{\mathrm{ o},l}^{2}} \right.\nonumber\\
&+&\left.\sum_{k}\sum_{i,j}[w^{\rm o}_{k}(\theta_{i})-w^{\rm m}_{k}(\theta_{i})](C_{k}^{-1})_{i,j}[w^{\rm o}_{k}(\theta_{j})-w^{\rm m}_{k}(\theta_{j})]\right] \nonumber\\ 
&+&\sum_{l}\ddfrac{\log(\Phi^{\rm o,sf}_{l}/\Phi^{\rm o,p}_{l}) - \log(\Phi^{\rm m,sf}_{l}/\Phi^{\rm m,p}_{l})}{\log\sigma_{\mathrm{ o,sf},l}^{2} + \log\sigma_{\mathrm{o,p},l}^{2}}\rm.
\end{eqnarray}

Here, the indices `p' and `sf' relate to our passive and \np\ populations, $l$ to bins of stellar mass, $i$ and $j$ to bins of angular separation and $k$ to stellar mass selections. The superscripts `o' and `m' refer to observational data and model prediction, respectively.  The first line relates to our stellar mass functions, the second to our two-point angular correlation functions and the third to the ratio of our stellar mass functions.
  
We obtain good results only using the two-point correlation function for $\log(M_{\star}/\mathrm{M}_{\odot})>9.8$ and $>10.6$ populations as constraints in our calibration. We have verified that including additional mass bins results in negligible changes to the resulting best-fit model, so we opt for a simpler model in that it has fewer observational constraints.  This choice also means that the correlation (two-point) and stellar mass function (one-point) data contribute roughly equal amounts to the overall $\chi^{2}$. We thus have $59$ constraining data points, which results in $59-15=44$ degrees of freedom ($N_{\rm dof}$). 

We use the affine-invariant {\sc python} implementation for Markov chain Monte Carlo (MCMC), {\sc emcee} \citep{Foreman-Mackey:2013} in order to fit our model. We employ 60 walkers for $10^{4}$ iterations after a burn-in phase of $2\times10^{3}$ and test for convergence using the $\hat{R}$ statistic \citep{GelmanRubin:1992}. Our one-dimensional and two-dimensional marginalised posterior distributions are shown in Appendix~\ref{sec:fitting_appendix}, along with our convergence tests. We assume flat priors for each free parameter. The best-fit values are taken to be the median of the posterior distribution, and the uncertainties the $16-84^{\rm th}$~percentiles. These are summarized in Table~\ref{tab:params} in Appendix~\ref{sec:fitting_appendix}. Our best-fit model achieves a reduced $\chi^{2}$ of $\chi^{2}/N_{\rm dof}\sim1.9$, indicating a reasonable fit to the data.

\section{Results}
\label{sec:results}
In this section, we present our main results. The performance of the best-fit model with respect to the calibration data (the observed stellar mass functions, their ratio, and the angular two-point correlation functions) is shown in Section~\ref{sec:best_fit_model}. In Section~\ref{sec:fshmr} we investigate the resulting SHMRs for central galaxies (see \autoref{eq:fshmr_epsilon}) and perform some further modeling of these with $\Lambda$CDM halo mass accretion histories. Satellite galaxies are explored in Section~\ref{sec:satellites}.     
\subsection{The Best-fit Model}
\label{sec:best_fit_model}
\begin{figure*}
\centering
\includegraphics[width=7.10140in]{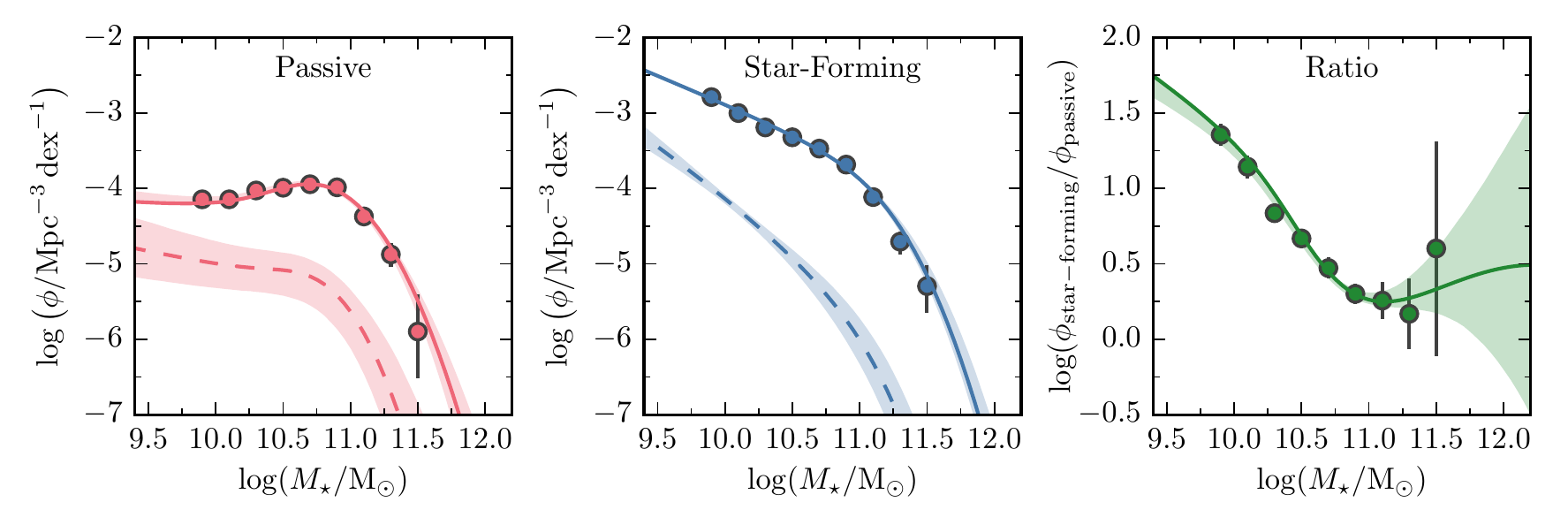}
\caption{The stellar mass functions for SMUVS galaxies for $2<z<3$ shown as points with ($1\sigma$) error bars for passive (left panel) and \np\ (middle panel) galaxies. The ratio of the stellar mass functions of the two populations is shown in the right panel. Our best-fit model is shown as the solid line with the $16-84^{\rm th}$~percentiles derived from our MCMC chains shown as the shaded region. The dashed line indicates the contribution to the stellar mass function from satellite galaxies, with the accompanying shaded region representing the $16-84^{\rm th}$~percentiles derived from our MCMC chains.}
\label{fig:best_gsmf}
\end{figure*}
\begin{figure*}
\centering
\includegraphics[width=7.10140in]{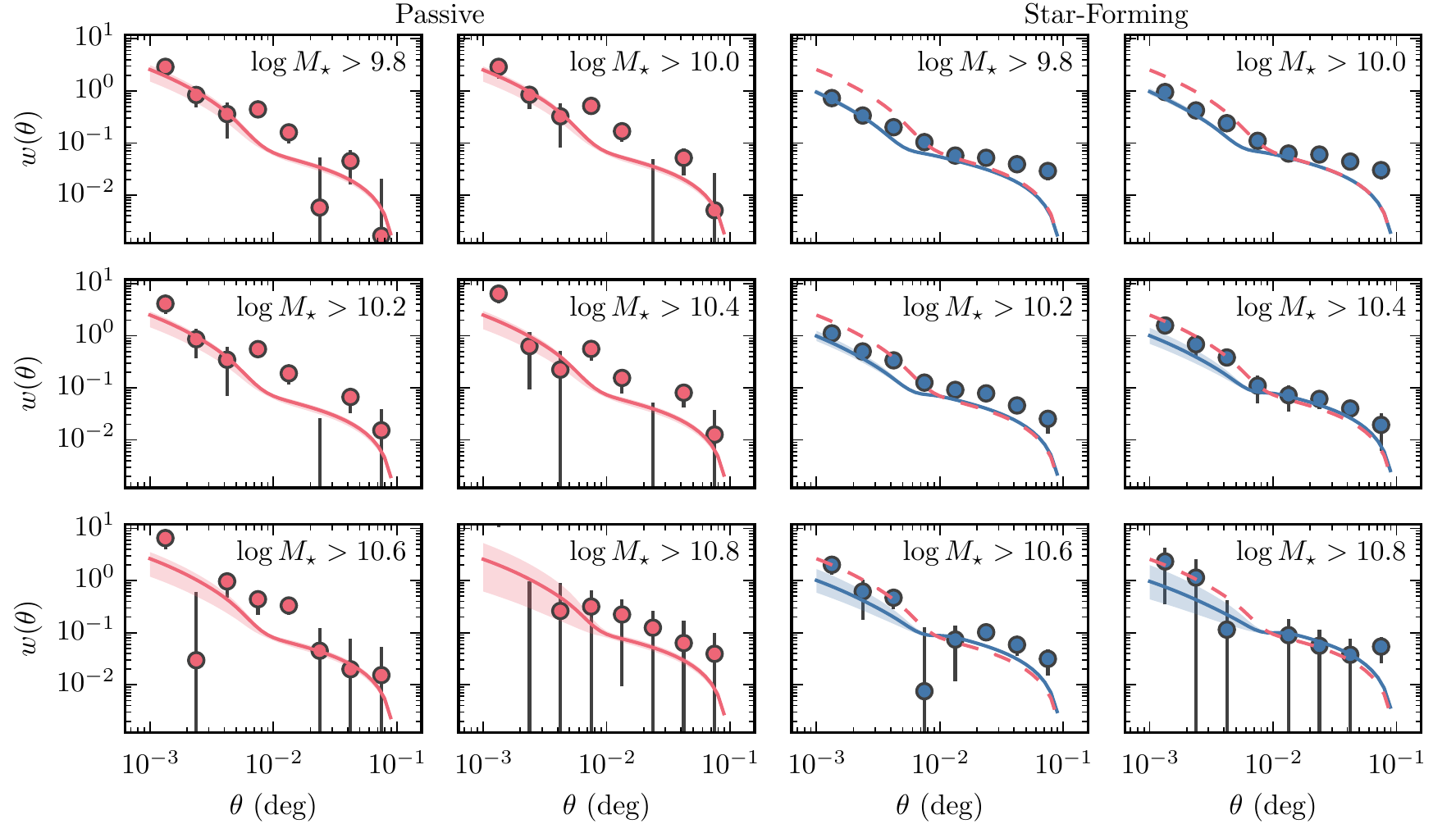}
\caption{The two-point angular correlation functions measured from SMUVS data (points with error bars) and the results from our best-fit model (solid lines). Passive (left two columns) and \np\ (right two columns) are indicated by red and blue colors respectively. For reference, the results of the best-fit model for passive galaxies are shown as the dashed-red lines in the \np\ panels. The shaded regions indicate the $16-84^{\rm th}$~percentiles derived from our MCMC chains. The stellar mass limit used is indicated in each panel, in M$_{\odot}$. Only the $\log(M_{\star}/\mathrm{M}_{\odot})>9.8$ and $>10.6$ populations are used as calibration data in our model fitting. Note that the models, rather than the observed data, have been corrected for the integral constraint, as discussed in Section~\ref{sec:project_photz}.}
\label{fig:best_wtheta}
\end{figure*}
In \autoref{fig:best_gsmf}, we show the observed galaxy stellar mass functions for our passive and \np\ populations, as well as the ratio between the two, as with the results from our best-fit model shown for comparison. The shape of the high-mass end of the stellar mass functions are substantively similar and the main difference between them is that the slope of the function for passive galaxies at masses smaller than the `knee' is much flatter. Both stellar mass functions can be accurately reproduced by the model.

In \autoref{fig:best_wtheta} we present the two-point angular correlation functions for passive and \np\ galaxies, with results from the best-fit model also shown. The large-scale clustering (i.e., two-halo term) of the two populations is similar. However, there is more power on small scales for the passive population, which suggests a larger fraction of these galaxies are satellites (we will explore this further in Section~\ref{sec:satellites}). Whilst passive galaxies are recognized as being more clustered in general at lower redshift \citep[e.g.][]{Zehavi:2011}, this appears to be less pronounced on larger scales and with increasing redshift \citep[e.g.][]{Hartley:2010}. We also note that \cite{Tinker:2013} find similar differences between their passive and \np\ correlation functions for $0.2<z<1.0$. Again, we can see that the model can reasonably reproduce the observed correlation functions, though there appear to be some mild systematic discrepancies at (i) $\sim10^{-2}$~deg for passive galaxies, which could be related to complex nonlinear biases (this scale approximately coincides with the transition from the one- to two-halo term) not properly accounted for in our model; and (ii) at large scales ($\gtrsim2\times10^{-2}$~deg) for star-forming galaxies with $\log(M_{\star}/\mathrm{M}_{\odot})\lesssim10.2$. Here, the discrepancy may be related to our decision to fix $\Delta_{z}$, which can control the amplitude of the correlation function independent of the halos the galaxies occupy \citep[e.g.][]{Cowley:2018}. We have investigated more complex models in which $\Delta_{z}$ was a free parameter for both passive and star-forming samples and had a stellar mass dependence. However, as mentioned earlier, we find that our current data are insufficient to robustly constrain these models, although this may help alleviate this mild discrepancy. 

In general, these discrepancies are fairly minor and the results shown here, combined with the reduced chi-squared value of $\chi^{2}/N_{\rm dof}\sim1.9$ mentioned earlier, indicate that our model can reproduce the calibration data to a good degree of accuracy.

\subsection{The Stellar-to-halo Mass Ratios}
\label{sec:fshmr}    
\begin{figure}
\centering
\includegraphics[width=3.35289in]{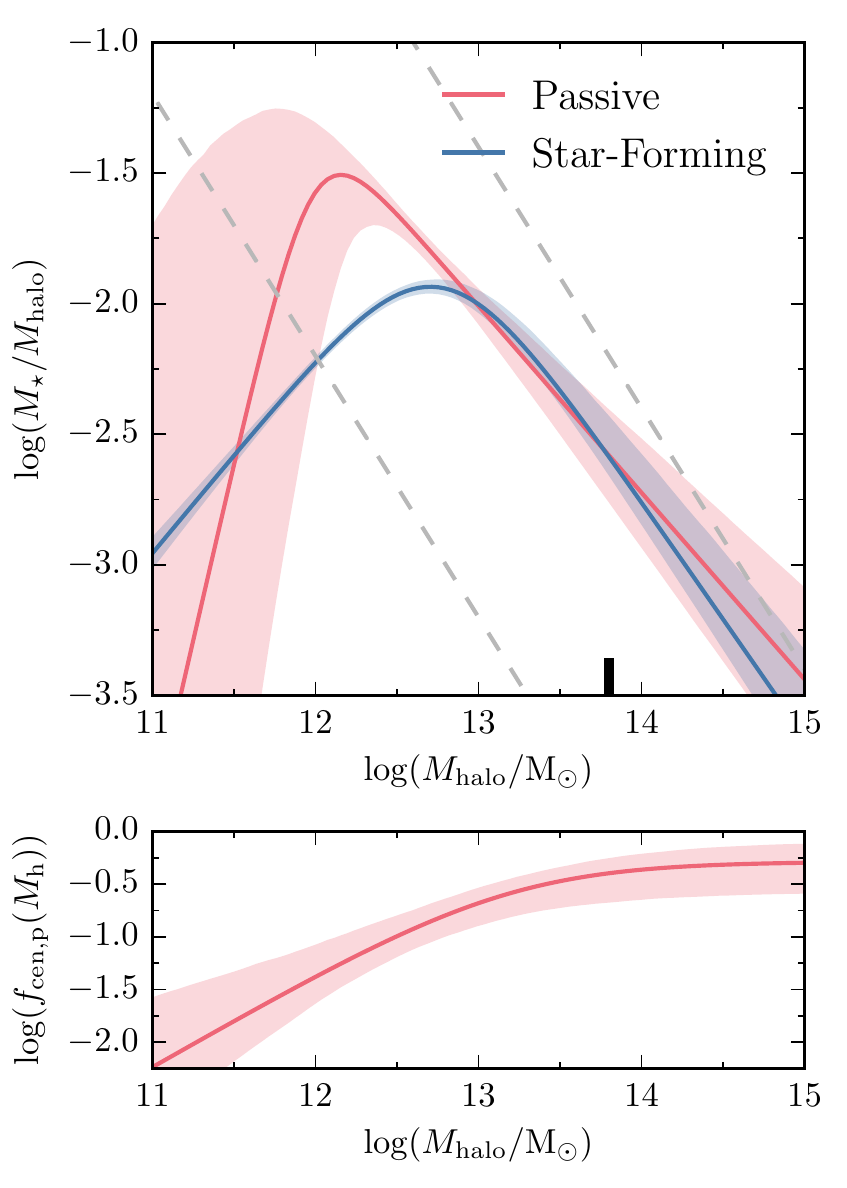}
\caption{Main panel: the stellar-to-halo mass ratios (\autoref{eq:fshmr_epsilon}) for central galaxies derived from our MCMC fitting for passive (red) and \np\ galaxies (blue). The shaded regions indicate the $16-84^{\rm th}$~percentiles of this relation derived from our MCMC chains. The dashed lines indicate lines of constant stellar mass at $\log(M_{\star}/\mathrm{M}_{\odot})=9.8$ and $11.6$ (i.e., the range of stellar masses constrained by our stellar mass functions). The thick black tick mark indicates the maximum halo mass expected in the SMUVS volume at $z\sim2-3$. Lower panel: the fraction of central galaxies that are passive as a function of halo mass (\autoref{eq:frac_cen_p}).}
\label{fig:fshmr}
\end{figure}

\begin{figure}
\centering
\includegraphics[width=3.35289in]{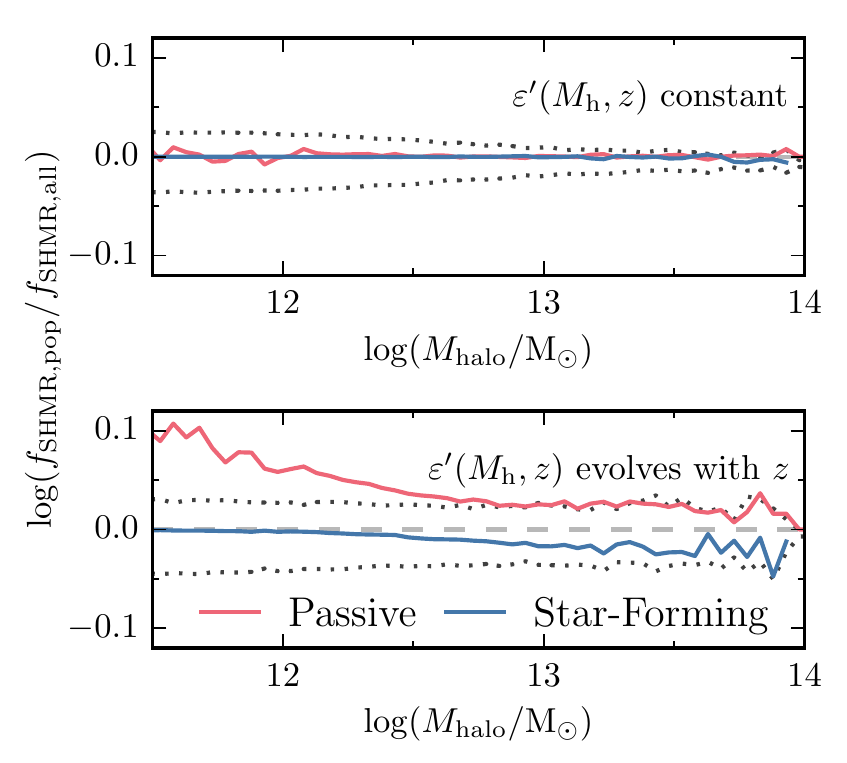}
\caption{Top panel: the stellar-to-halo mass ratios for passive (red) and \np\ galaxies (blue) divided by the stellar-to-halo mass ratio for the combined population. The $16-84^{\rm th}$~percentile scatter for the combined population is shown by the dotted black lines. A dashed gray line at unity is shown for reference. Bottom panel: as for the middle panel but for the evolving $\epsilon_{0}$ model.}
\label{fig:fshmr_halo_z}
\end{figure}
\begin{figure}
\centering
\includegraphics[width=3.35289in]{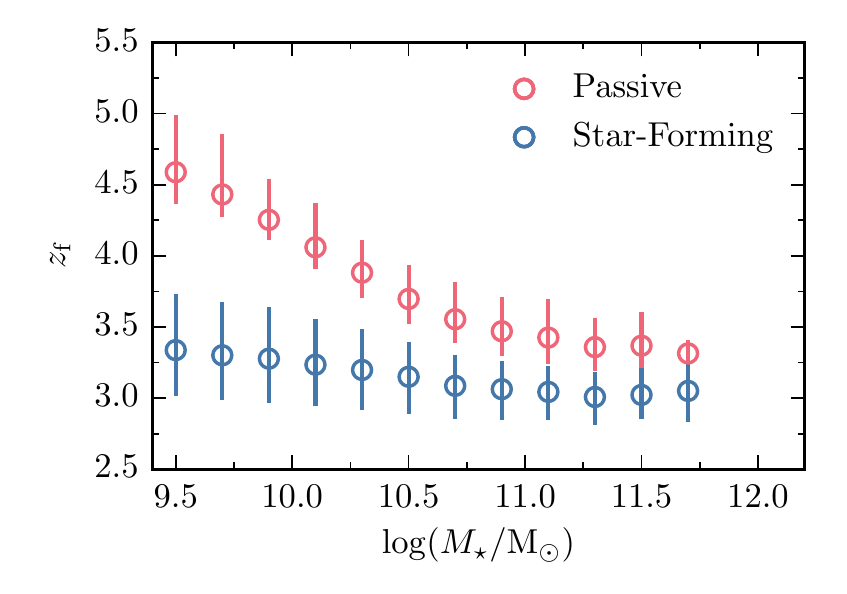}
\caption{The halo formation redshifts (defined as the redshift at which that halo had acquired half of its $z=2.3$ mass) of our passive and \np\ populations (red and blue symbols, respectively) for the evolving $\varepsilon(M_{\rm h},z)$ model. The open circles indicate the median formation time at a given stellar mass and the error bars the $16-84^{\rm th}$~percentile scatter.}
\label{fig:gsmf_halo_z}
\end{figure}

\autoref{fig:fshmr} shows the best-fit SHMR, $\varepsilon(M_{\rm h})$ (\autoref{eq:fshmr_epsilon}), for passive and \np\ central galaxies and the fraction of central galaxies that are passive as a function of host halo mass (\autoref{eq:frac_cen_p}). The high-mass slopes of the ratios are similar for both populations, as would be expected from the similarity in the high-mass regime of the stellar mass functions. Interestingly, the \np\ population relation peaks at a larger halo mass (by $\sim0.5$~dex) than the passive one (though at a similar stellar mass, $\sim10^{10.7}$~M$_{\odot}$, roughly coinciding with the knee of the observed stellar mass functions) and the passive relation has a higher normalization for $M_{\rm h}\lesssim10^{13}$~M$_{\odot}$.  We now investigate this difference in normalization further using a simple model comprising a realistic ensemble of halo histories.

\cite{McBride:2009} found that the mass assembly histories of dark matter halos in the Millennium $N$-body simulations \citep{Springel:2005} could be well described with the two parameter functional form
\begin{equation}
M(z) = M_{0}\,(1+z)^{\beta_{\rm h}}\,e^{-\gamma_{\rm h} z}\rm,
\end{equation}
where $M_{0}$ is the halo mass at $z=0$. We generate a realistic ensemble of $\beta_{\rm h}$ and $\gamma_{\rm h}$ values according to Appendix A of McBride et al. and weight the halos such that our halo mass function at $z=2.3$ is reproduced. From this, we generate the star formation histories, $\psi(M_{\rm h},z)$, in each halo according to 
\begin{equation}
\psi(M_{\rm h},z) = \varepsilon^{\prime}(M_{\rm h},z) \, f_{\rm b} \, \dot{M}_{\rm h}
\end{equation} 
\citep[e.g.][]{Tacchella:2018} where $f_{\rm b}=\Omega_{\rm b}/\Omega_{\rm m}$ is the baryon fraction and $\varepsilon^{\prime}(M_{\rm h},z)$ has the same form as \autoref{eq:fshmr_epsilon}, though has a different meaning than in \autoref{eq:fshmr} so is denoted with the `prime' symbol here for clarity, as are its parameters.  Additionally, we compute the stellar mass of each galaxy, assuming the instantaneous recycling approximation, through
\begin{equation}
M_{\star} = (1-R)\int\,\psi(M_{\rm h},z)\,\mathrm{d}t\rm,
\end{equation} 
where $R$ is the returned fraction, the fraction of initial stellar mass that is returned to the interstellar medium by mass loss from dying stars. We use a value of $R=0.41$, which corresponds to our choice of a \cite{Chabrier:2003} IMF. For computing stellar mass functions, we add a Gaussian scatter with $\sigma_{\log M_{\star,\rm meas}}=0.13$~dex onto our intrinsic stellar masses, as we did in Section~\ref{sec:model}. In bins of stellar mass, we then label galaxies with the lowest intrinsic specific star formation rates as passive such that the difference between the passive and \np\ central galaxy stellar mass functions is reproduced. We assume no redshift evolution in $\varepsilon^{\prime}(M_{\rm h},z)$ and find that the parameter values ($\epsilon_{0}^{\prime}$, $M_{\rm c}^{\prime}$, $\beta^{\prime}$, $\gamma^{\prime}$) = ($0.15$, $2.1\times10^{12}$~$h^{-1}$~M$_{\odot}$, $0.8$, $0.8$) give reasonable results, i.e., can reproduce our observed stellar mass function. 

The SHMRs for passive and \np\ galaxies from this model are shown in the top panel of \autoref{fig:fshmr_halo_z}. They are essentially identical, implying that for the difference in normalization we found in \autoref{fig:fshmr} to arise $\varepsilon^{\prime}(M_{\rm h},z)$ \emph{must} evolve with redshift. Indeed, we illustrate this further by repeating the above exercise but now allowing $\epsilon_{0}^{\prime}$ to vary as a function of redshift. For simplicity we choose $\epsilon_{0}^{\prime}(z)=\epsilon_{\rm N}^{\prime} + \epsilon_{z}^{\prime}\times(1+z)$ and values of ($\epsilon_{\rm N}^{\prime}$, $\epsilon_{z}^{\prime}$) = ($0.04$, $0.03$) such that the conversion of baryons into stars becomes more efficient at higher redshifts (and our observed stellar mass functions are still reasonably reproduced), all other parameters are kept the same. The results from this model are shown in the bottom panel of \autoref{fig:fshmr_halo_z}. As expected, we find that the normalization of the SHMR for passive galaxies is elevated relative to that of the star-forming galaxies. We note that we have not precisely reproduced the differences found in \autoref{fig:fshmr}, as the evolution of $\varepsilon^{\prime}(M_{\rm h},z)$ is likely more complex than investigated here, but we have illustrated how such differences can occur.

For each halo history, we now calculate a halo formation redshift, $z_{\rm f}$, which we define as the redshift at which that halo had assembled half of its $z=2.3$ mass. We show these for passive and \np\ galaxies from the evolving $\varepsilon^{\prime}(M_{\rm h},z)$ model in \autoref{fig:gsmf_halo_z}, though we note that this Figure would be substantively similar if the constant $\varepsilon^{\prime}(M_{\rm h},z)$ model were considered. Interestingly, we find that selecting passive galaxies by choosing those with the lowest specific star formation rates results in us selecting halos with the highest formation redshifts, i.e., those that are assembling mass at the lowest rates.\footnote{We note that \cite{Feldmann:2016} found a similar result studying passive galaxies in the FIRE simulations (see also \citealt{Feldmann:2015}).} This helps further explain the result shown in \autoref{fig:fshmr_halo_z}. Passive galaxies tend to reside in `older' halos, thus gaining a larger proportion of their final stellar mass at higher redshifts when the conversion from baryons into stars was more efficient. This results in a higher SHMR, such as that found in \autoref{fig:fshmr}. 

Additionally, on further inspection of \autoref{fig:gsmf_halo_z}, we see that the halo formation redshift for \np\ galaxies is not strongly dependent on stellar mass but for passive galaxies, there is a much stronger trend that flattens for $M_{\star}\gtrsim10^{10.75}$~M$_{\odot}$. This could explain the difference in the evolution of the number density of intermediate- and high-mass passive galaxies found by \cite{Deshmukh:2018}, as these stellar mass ranges appear to select halos with very different rates of growth.         
    
\subsection{Satellite Fractions and Quenching}
\label{sec:satellites}
\begin{figure*}
\centering
\includegraphics[width=7.10140in]{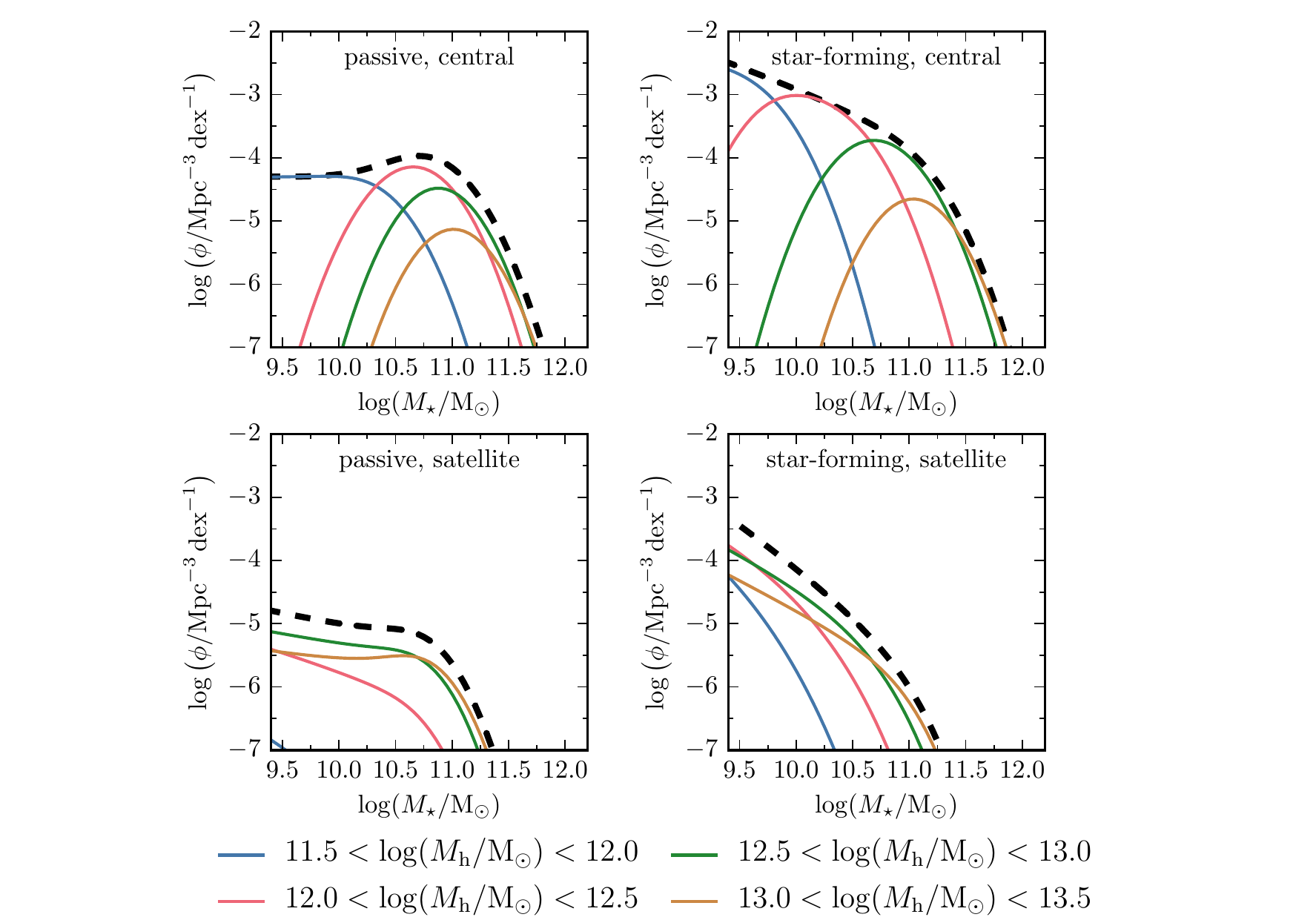}
\caption{Galaxy stellar mass functions of passive (left panels) and \np\ (right panels) galaxies, divided into central (top panels) and satellite galaxies (right panels), as predicted by our best-fit model. In each panel, the total stellar mass function is shown as a dashed black line and the contribution from different halo masses as colored lines, as indicated in the legend.}
\label{fig:cond_gsmf}
\end{figure*}
\begin{figure}
\centering
\includegraphics[width=3.35289in]{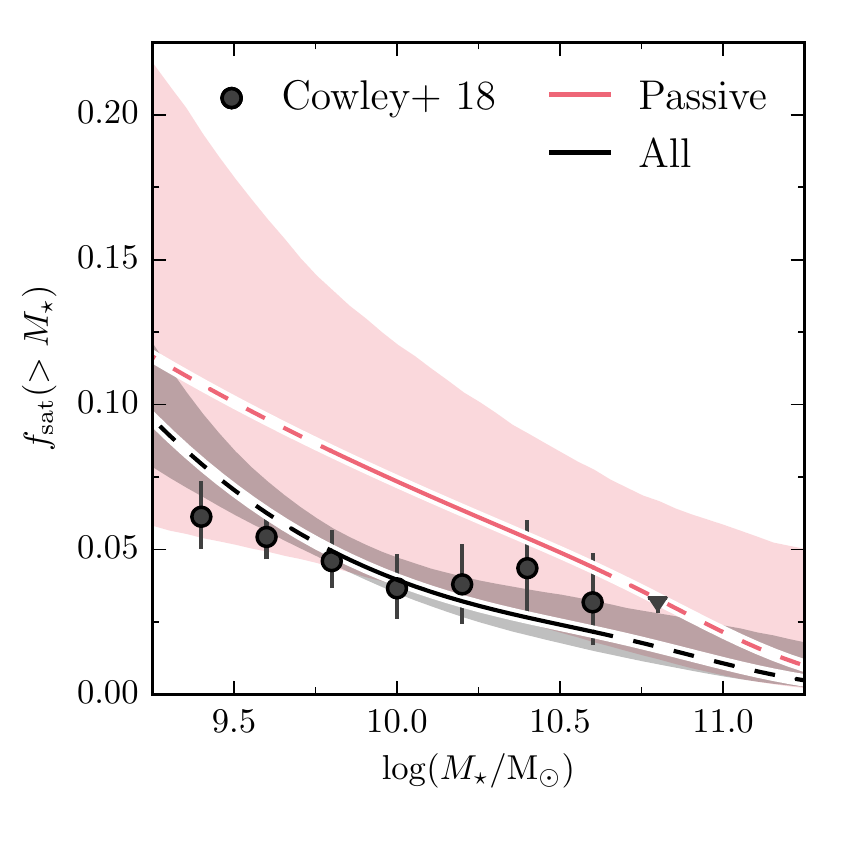}
\caption{The cumulative satellite fraction as a function of stellar mass for passive and all galaxies (red and black solid lines, respectively, the dashed portion of the lines indicates stellar masses for which our model is not directly constrained by clustering data). The shaded regions indicate the $16-84^{\rm th}$~percentiles derived from our MCMC chains. Observational data (for all galaxies) are from Cowley et al. (\citeyear{Cowley:2018}, black circles).}
\label{fig:fsat}
\end{figure}

In this section, we address the abundance of satellites in our galaxy populations and discuss potential physical quenching mechanisms for them. 

As a result of our halo formalism, we can compute the contribution to the stellar mass function from a given halo mass by appropriately changing the integration limits in \autoref{eq:cond_gsmf}. We show the contribution to the stellar mass function from a selection of halo masses and divided into passive/\np\ and central/satellite in \autoref{fig:cond_gsmf}. Here we can see that at a fixed stellar mass the satellite mass function is dominated by greater mass halos than the central one, and this is especially pronounced for passive galaxies. The stellar mass of central galaxies forms a relatively tight relationship with the mass of their host halos (e.g., \autoref{eq:fshmr_epsilon}) but this becomes offset once they are accreted into a more massive host halo and become a satellite. Their stellar mass will not necessarily change much following the halo merger but their host halo mass will suddenly have increased, so this result is unsurprising. Additionally, we can see that the contributions from each halo mass range form a fairly well-defined peaked function for central galaxies, but this is not the case for satellites, indicating that a relationship between the stellar mass of a satellite galaxy and the mass of its host halo must have a much greater scatter than for centrals. This can be understood as the merger event and subsequent environmental effects scattering satellite galaxies from the initial (relatively tight) SHMR they had as central galaxies.   

We show the satellite fractions for our passive population and the complete mass selected population (passive + star-forming) in \autoref{fig:fsat}. We see that our results for the complete population compare favorably with our earlier analysis in \cite{Cowley:2018}, which used a less-sophisticated halo occupation model than we have adopted here. We can also see that the satellite fractions for passive galaxies are generally greater (as has been found at lower redshifts, e.g., \citealt{Knobel:2013}), though the difference is fairly small and there are significant uncertainties on the satellite fraction for passive galaxies. This is unsurprising, as this quantity is constrained mainly by the small-scale clustering data for passive galaxies that have significant uncertainties. However, a greater satellite fraction indicates that a passive galaxy is more likely to be a satellite than one selected from the combined population, and implies that the environmental effects satellite galaxies are subject to can play a role in quenching ongoing star formation. We explore this idea in greater detail below.     

\subsubsection{The Astrophysics of Satellite Quenching}
\begin{figure}
\centering
\includegraphics[width=3.35289in]{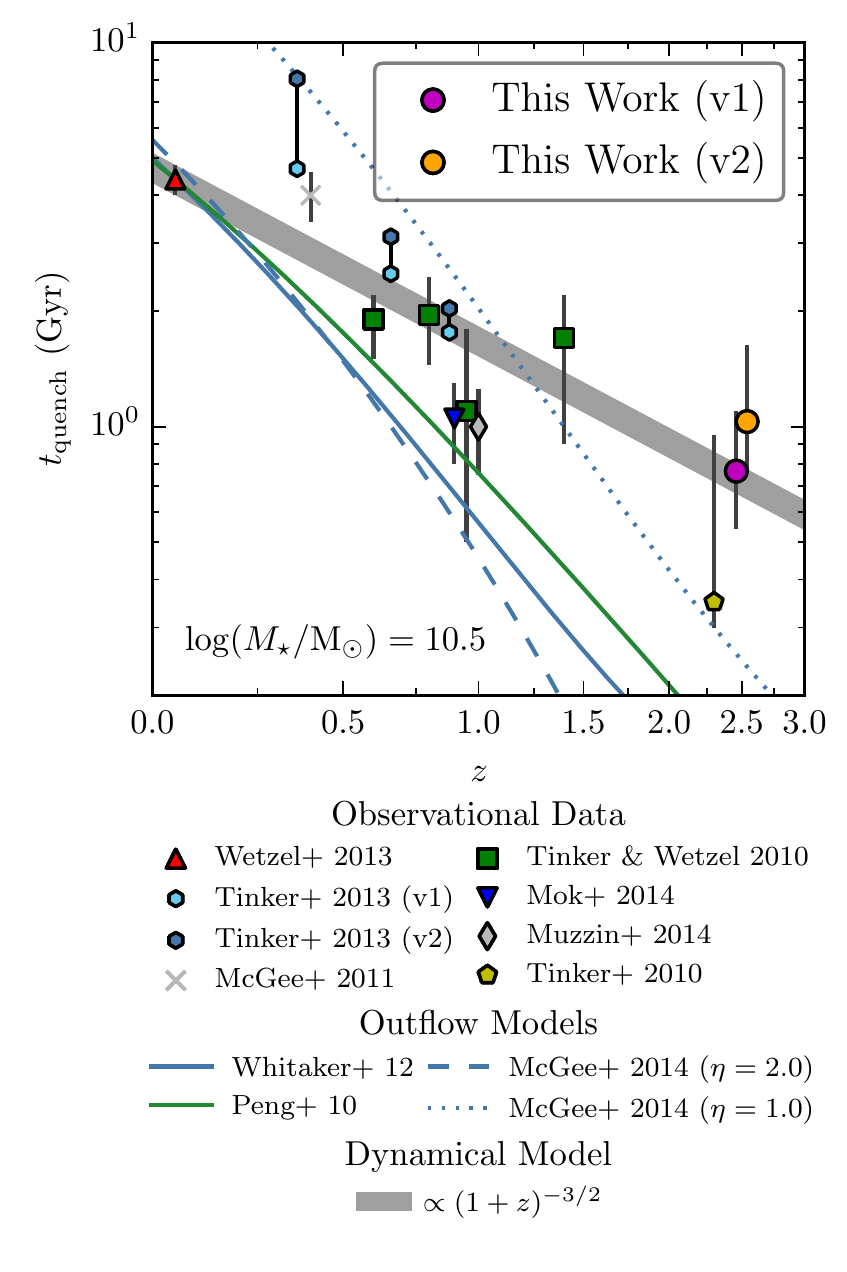}
\caption{The quenching timescale of satellite galaxies with $\log(M_{\star}/\mathrm{M}_{\odot})=10.5$. Observational data are from Wetzel et al. (\citeyear{Wetzel:2013}, red triangle), Tinker et al. (\citeyear{Tinker:2013}, hexagons), McGee et al. (\citeyear{McGee:2011}, gray cross), Mok et al. (\citeyear{Mok:2014}, blue downward triangle), Muzzin et al. (\citeyear{Muzzin:2014}, gray diamond) and Tinker et al. (\citeyear{TinkerWechslerZheng:2010}, yellow pentagon). Outflow models based on the galaxy `main-sequence' star formation rate parameterizations of Whitaker et al. (\citeyear{Whitaker:2012}, blue line) - with the modifications of McGee et al. (\citeyear{McGee:2014}) for a mass-loading of $\eta=1$ and $2$ (blue dotted and dashed lines respectively) - and Peng et al. (\citeyear{Peng:2010}, green line). A dynamical model (scaled to the $z=0$ estimate of Wetzel et al.) is shown by the gray shaded region. Estimates from this work assume either no galaxies were quenched upon infall (v1, purple point) or that the passive fraction of central galaxies were quenched (v2, orange point).}
\label{fig:tquench_z}
\end{figure}
In this section, we infer satellite quenching timescales and compare these with other estimates from the literature at lower redshift, to briefly discuss the possible physical mechanisms for satellite quenching, i.e., dynamical/orbit- or outflow-based, which can generally be recognized as `environmental' and `intrinsic/mass-related' respectively. In doing so, we are broadly revisiting the discussion of \cite{McGee:2014}, though arrive at a different conclusion.

We derive a quenching timescale for our satellite population based on the method of \cite{TinkerWetzel:2010} and \cite{Tinker:2013}. At each stellar mass, we compute the fraction of satellite galaxies that are passive, and then compare this to the age distribution of satellite halos (i.e., the time since they were last a distinct halo) at this redshift taken from the recent $P$-Millennium simulation \citep{Baugh:2019}. Using the ansatz that the quenched (i.e., passive) galaxies are the ones that have been satellites for the longest period of time, we can then derive a quenching timescale. For example, if $50$~\% of satellite galaxies are passive and $50$~\% of satellite-halos have been a satellite for $1$~Gyr or more then our quenching timescale would be $1$~Gyr. However, some satellites may have been passive prior to accretion onto their current host. To account for this, we follow \cite{Tinker:2013} and assume two extremes: either that no galaxies were passive prior to being accreted (v1) or that the fraction of central galaxies that are passive at $z=2.3$ was already passive\footnote{In reality, this fraction will be nonzero but less than it is at $z=2.3$ because galaxies were accreted at higher redshifts when the fraction of central galaxies that are passive will have been lower.} (v2). For example, if $50$~\% of satellite galaxies are passive but $10$~\% of central galaxies are passive, then the quenching timescale we are concerned with in v2 will be the $40^{\rm th}$~percentile of the satellite age distribution, whereas in v1 it would be the $50^{\rm th}$~percentile.

In the case that quenching depends only on the orbit of the satellite within its host halo then the quenching timescale will evolve as the ratio of the inverse densities, $\propto (1+z)^{-3/2}$ [for a halo in virial equilibrium the dynamical time scales as $\rho^{-1/2}$ and the density, $\rho$, scales as $(1+z)^{3}$]. However, if supernovae-driven outflows are primarily responsible then the quenching time will scale more strongly with the star formation rate. McGee et al. used the fitted prescriptions for the star-forming `main sequence' of \cite{Peng:2010} and \cite{Whitaker:2012} to investigate the potential evolution of the quenching timescale under this scenario, assuming that the outflow rate is invariant with redshift for a given star formation rate and stellar mass. They also modified the prescription of Whitaker et al. to account for different mass-loading factors, $\eta$, where a larger value means more cold gas mass is ejected from a galaxy per unit star formation, thus inhibiting subsequent star formation. We show the scaling of these outflow models [normalized to the quenching timescale of \cite{Wetzel:2013} for $M_{\star}=10^{10.5}$~M$_{\odot}$ galaxies at $z\sim0$] in \autoref{fig:tquench_z}. We also include in this Figure a broad range of other estimates of the quenching timescale of $10^{10.5}$~M$_{\odot}$ satellite galaxies, including our own. At $M_{\star}=10^{10.5}$~M$_{\odot}$ we find $t_{\rm quench}=0.77^{+0.33}_{-0.22}$~Gyr for v1 and $1.03^{+0.60}_{-0.27}$~Gyr for v2. 

The various observational data seem to cover a plethora of possible scenarios, most likely related to the use of different techniques and methods to derive a quenching timescale. This highlights the need for a homogeneous analysis to be performed over a broad range of redshifts and means that we are not able to draw strong conclusions here. McGee et al. considered only the data of \cite{Wetzel:2013}, \cite{McGee:2011}, \cite{Mok:2014} and \cite{Muzzin:2014} and so arrived at the conclusion that outflow-based models were in better agreement with the data. In contrast, our estimates, taken in conjunction with those of \cite{Wetzel:2013}, seem to strongly favor a dynamical mechanism for satellite quenching, e.g., ram pressure stripping. This is the same conclusion that \cite{TinkerWetzel:2010} arrived at and it appears from inspection of \autoref{fig:tquench_z} that more high-redshift data, such as that presented in this work, could help further discriminate between these two potential processes. 

\section{Summary}
\label{sec:conclude}
We have used a sophisticated halo occupation model to investigate the stellar-to-halo mass ratios (SHMRs) of passive and \np\ galaxies, as identified in the SMUVS survey, at $z\sim2-3$. The free parameters in the model are calibrated through fitting them to the observed galaxy stellar mass and angular two-point correlation functions for each population. The model provides a meaningful way to statistically assign an average halo mass to an observed stellar mass and to divide our galaxy populations into central and satellite galaxies.

We find that the central galaxy SHMRs are different for passive and \np\ galaxies. Interestingly, the normalization of this relation around its peak is higher for passive galaxies than that of the \np\ population. Using a simple model based on $\Lambda$CDM halo mass accretion histories, we show that this can arise if the efficiency with which baryons are converted into stars at a given halo mass evolves with redshift, such that it is more efficient at higher redshifts. This mass accretion model also shows that passive central galaxies can be plausibly explained as simply a selection of the host halos with the highest formation redshifts, i.e., those with the lowest mass accretion rates. The halo formation redshift is very dependent on stellar mass for passive galaxies and this may help to explain the very different evolution of high- and intermediate-stellar mass selected passive galaxies found by \cite{Deshmukh:2018}.   

At a given stellar mass, satellite galaxies occupy host halos of a greater mass than their central counterparts. Additionally, the satellite fraction of passive galaxies is greater than that of the combined population (passive and \np) indicating that environmental processes are expected to play a role in galaxy quenching. 

Assuming that passive (i.e., quenched) satellites will be those that have been satellites for the longest and using the distribution of satellite (sub-halo) ages from an $N$-body simulation we derive satellite quenching timescales. Through comparing our estimates, and others from the literature at lower redshifts, to the results from very simple dynamical- and outflow-based models, we find that our data seem to favor a dynamically driven quenching model, e.g., ram pressure stripping. However, it is possible to draw different conclusions depending on the observational data considered, as they appear to support a range of possible scenarios. It appears that further studies of the satellite quenching timescale at higher redshifts will allow for a more definitive answer to this particular question.
  
\section*{Acknowledgements}
The authors would like to thank the anonymous referee for a comprehensive and constructive report that allowed us to improve the overall quality of the manuscript. This work is based in part on observations carried out with the \emph{Spitzer Space Telescope}, which is operated by the Jet Propulsion Laboratory, California Institute of Technology under a contract with NASA. It also uses data products from observations conducted with ESO Telescopes at the Paranal Observatory under ESO program ID 179.A-2005, data products produced by TERAPIX and the Cambridge  Astronomy Survey Unit on behalf of the UltraVISTA consortium, observations carried out by NASA/ESA \emph{Hubble Space Telescope} obtained and archived at the Space Telescope Science Institute; and the Subaru  Telescope, which is operated by the National Astronomical Observatory of Japan. This research has made use of the NASA/IPAC Infrared Science Archive, which is operated by the Jet Propulsion Laboratory, California Institute of Technology, under contract with NASA. 

This work has made use of the following open-source \textsc{python} packages: \textsc{numpy} \citep{numpy}, \textsc{scipy} \citep{scipy}, \textsc{matplotlib} \citep{Hunter:2007} and \textsc{ipython} \citep{ipython}. The color-vision-impaired-friendly color schemes used throughout can be found at \url{https://personal.sron.nl/~pault/}. 

W.I.C., S.D., and K.I.C. acknowledge funding from the European Research Council through the award of the Consolidator grant ID 681627-BUILDUP. The Cosmic Dawn Center is funded by the Danish National Research Foundation.
\appendix
\section{MCMC Fitting}
\label{sec:fitting_appendix}
\begin{figure*}
\centering
\includegraphics[width=7.10140in]{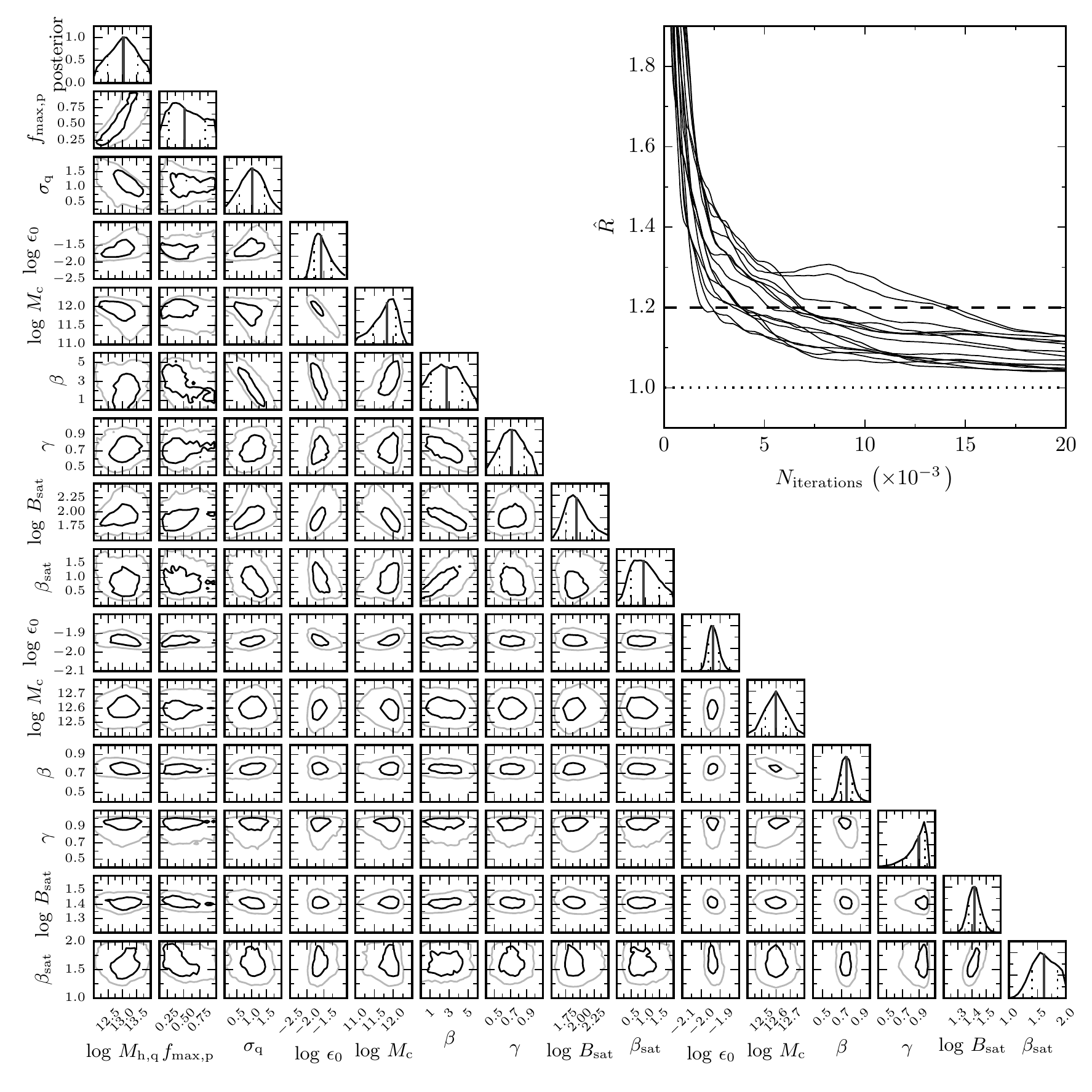}
\caption{Top right panel: the Gelman \& Rubin (\citeyear{GelmanRubin:1992}) $\hat{R}$ statistic for each of our MCMC parameters (each line represents a parameter) at each iteration in the MCMC. Reference values of $1.2$ and $1.0$ are shown as the black dashed and dotted lines, respectively. A value of $\hat{R}<1.2$ indicates convergence. Other panels: the marginalized 1D (diagonal panels) and 2D (off-diagonal) posterior distributions for our MCMC parameters. The solid vertical lines in the diagonal panels indicate the median of the posterior distribution (taken to be our best-fit value) and the dashed lines indicate the $16-84^{\rm th}$~percentiles. The black and gray contours in the off-diagonal panels indicate the $1\sigma$ and $2\sigma$ levels respectively. The rightmost six parameters are for \np\ galaxies. This Figure was produced using code adapted from \cite{corner}. Halo masses are in $h^{-1}$~M$_{\odot}$. The limits of the axes are either less than or equal to the (flat) prior range used.}
\label{fig:corner}
\end{figure*}
\begin{deluxetable*}{lccc}
\tablecaption{Parameter values from MCMC fitting}
\tablecolumns{4}
\tablenum{1}
\tablewidth{0pt}
\tablehead{
\colhead{Parameter} &
\colhead{Value} &
\colhead{Prior Range} & 
\colhead{Eqn.}
}
\startdata
\multicolumn{4}{l}{Passive fraction:}\\ 
\rule{0pt}{3ex} $\log(M_{\mathrm{h,q}})$&$13.03^{+ 0.51}_{- 0.55}$&$(12.00,14.00)$&\ref{eq:frac_cen_p}\\  
\rule{0pt}{3ex} $f_{\mathrm{max,p}}$&$ 0.52^{+ 0.31}_{- 0.24}$&$( 0.00, 1.00)$&\\  
\rule{0pt}{3ex} $\sigma_{\mathrm{q}}$&$ 1.02^{+ 0.41}_{- 0.43}$&$( 0.10, 2.00)$&\\ 
\multicolumn{4}{l}{Passive Galaxies:}\\ 
\rule{0pt}{3ex} $\log(\epsilon_{0})$&$-1.59^{+ 0.30}_{- 0.21}$&$(-2.50,-0.82)$&\ref{eq:fshmr_epsilon}\\  
\rule{0pt}{3ex} $\log(M_{\mathrm{c}})$&$11.83^{+ 0.24}_{- 0.37}$&$(11.00,12.50)$&\\  
\rule{0pt}{3ex} $\beta$&$ 2.72^{+ 1.69}_{- 1.64}$&$( 0.00, 6.00)$&\\  
\rule{0pt}{3ex} $\gamma$&$ 0.71^{+ 0.15}_{- 0.15}$&$( 0.40, 1.10)$&\\  
\rule{0pt}{3ex} $\log(B_{\mathrm{sat}})$&$ 1.94^{+ 0.28}_{- 0.19}$&$( 1.50, 3.00)$&\ref{eq:Msat}\\  
\rule{0pt}{3ex} $\beta_{\mathrm{sat}}$&$ 0.95^{+ 0.54}_{- 0.44}$&$( 0.00, 2.00)$&\\  
\multicolumn{4}{l}{Star-Forming Galaxies:}\\ 
\rule{0pt}{3ex} $\log(\epsilon_{0})$&$-1.94^{+ 0.03}_{- 0.03}$&$(-2.50,-0.82)$&\ref{eq:fshmr_epsilon}\\  
\rule{0pt}{3ex} $\log(M_{\mathrm{c}})$&$12.60^{+ 0.07}_{- 0.07}$&$(11.50,12.80)$&\\  
\rule{0pt}{3ex} $\beta$&$ 0.75^{+ 0.06}_{- 0.06}$&$( 0.00, 2.50)$&\\  
\rule{0pt}{3ex} $\gamma$&$ 0.90^{+ 0.08}_{- 0.15}$&$( 0.40, 1.10)$&\\  
\rule{0pt}{3ex} $\log(B_{\mathrm{sat}})$&$ 1.42^{+ 0.04}_{- 0.04}$&$( 0.50, 2.50)$&\ref{eq:Msat}\\  
\rule{0pt}{3ex} $\beta_{\mathrm{sat}}$&$ 1.62^{+ 0.24}_{- 0.21}$&$( 0.50, 2.00)$&\\
\multicolumn{4}{l}{Fixed Parameters:}\\
\rule{0pt}{3ex} $\alpha_{\rm sat}$&$1.0$&--&\ref{eq:nsat}\\
\rule{0pt}{3ex} $\sigma_{\log(M_{\star})}$&$0.2$~dex&--&--\\
\rule{0pt}{3ex} $\Delta_{z}$&$0.035$&--&\ref{eq:sigma_z_window}
\enddata
\tablecomments{All halo masses are in $h^{-1}$~M$_{\odot}$.}\label{tab:params}
\end{deluxetable*}

In this Appendix, we give some brief details regarding the results of our MCMC fitting procedure. We use the \cite{GelmanRubin:1992} $\hat{R}$ statistic for each free parameter to assess whether our MCMC chains have converged. These are shown in the top right panel of \autoref{fig:corner}. All parameters have achieved values of $\hat{R}<1.2$, indicating that they have converged.

The marginalized $1$D and $2$D posterior distributions for our free parameters are shown in \autoref{fig:corner}. Whilst there are some degeneracies between parameters, this is not often the case, and the majority of the $1$D posterior distributions form well-defined peaks. Our best-fit parameter values and $16-84^{\rm th}$~percentile ranges from our MCMC chains are given in Table~\ref{tab:params}.
   
\bibliographystyle{aasjournal}
\bibliography{ref}
\end{document}